\documentclass[conference]{IEEEtran}
\IEEEoverridecommandlockouts
\usepackage{cite}
\usepackage{amsmath,amssymb,amsfonts}
\usepackage{algorithmic}
\usepackage{graphicx}
\usepackage{textcomp}
\usepackage{xcolor}
\usepackage{caption}
\usepackage{subcaption}

\def\BibTeX{{\rm B\kern-.05em{\sc i\kern-.025em b}\kern-.08em
    T\kern-.1667em\lower.7ex\hbox{E}\kern-.125emX}}
\graphicspath{ {figures/} }
\begin{document}

\title{Capacitive Sensor Based 2D Subsurface Imaging Technology for Non Destructive Evaluation of Building Surfaces }  

\author{\IEEEauthorblockN{Lasitha Piyathilaka$^{*}$, Basura Sooriyaarachchi, Sarath Kodagoda, Karthick Thiyagarajan}

\IEEEauthorblockA{
\textit{iPipes Lab, Centre for Autonomous Systems, School of Mechanical and Mechatronic Engineering}\\
\textit{University of Technology Sydney} \\
Sydney, Australia \\
$^{*}$Lasitha.Piyathilaka@uts.edu.au}

}

\maketitle


\begin{abstract}
Understanding the underlying structure of building surfaces like walls and floors is essential when carrying out building maintenance and modification work. To facilitate such work, this paper introduces a   capacitive sensor based technology which can conduct  non-destructive evaluation of building surfaces. The novelty of this sensor is that it can generate a  real-time 2D subsurface image which can be used to understand structure beneath the top surface. Finite Element Analysis (FEA) simulations are done to understand the best sensor head configuration that gives optimum results. Hardware and software components are custom-built to facilitate real-time imaging capability. The sensor is validated by laboratory tests, which revealed the ability of the proposed capacitive sensing technology to see through common building materials like wood and concrete. The 2D  image generated by the sensor is found to be useful in understanding the subsurface structure beneath the top surface. 
\end{abstract}


\section{Introduction}

Building surfaces like walls and floors undergo frequent modifications, rehabilitation, and restorations. It is important those work are carried out without affecting the underline building structure in the safest possible manner. Therefore, seeing through building surfaces is essential to detect studs, gas pipes, water pipes and electric wiring. If the subsurface of the building surfaces can be imaged in a non-destructive manner then the accidental damage to the underline structure of the building can be avoided. Most importantly health hazards due to electric shocks and gas leaks can be avoided.

Several Non Destructive Evaluation (NDE) sensors are used by the construction industry to see through walls. The stud detectors are the most common but they lack the 2D imaging capability. To understand the structure of the subsurface, several scans are done on a predefined pattern while making marks on the surface, which is a time-consuming operation and often not accurate.  Cover-meters are widely used to locate re-bars on concrete surfaces but only capable of detecting metal-based structures \cite{ulapane2020pulsed}. Radar-based technologies like Ground Penetrating Radar (GPR) are used to see through surfaces but they are not sensitive to near-surface variations due to the adversarial effect of the source wave \cite{ulapane2019some}. GPR scans are difficult to interpret by an inexperienced person and often post-processing of the raw data is required to get a clear understanding of the sub-surface.

\begin{figure}
    \centering
    \includegraphics[width=0.4\textwidth]{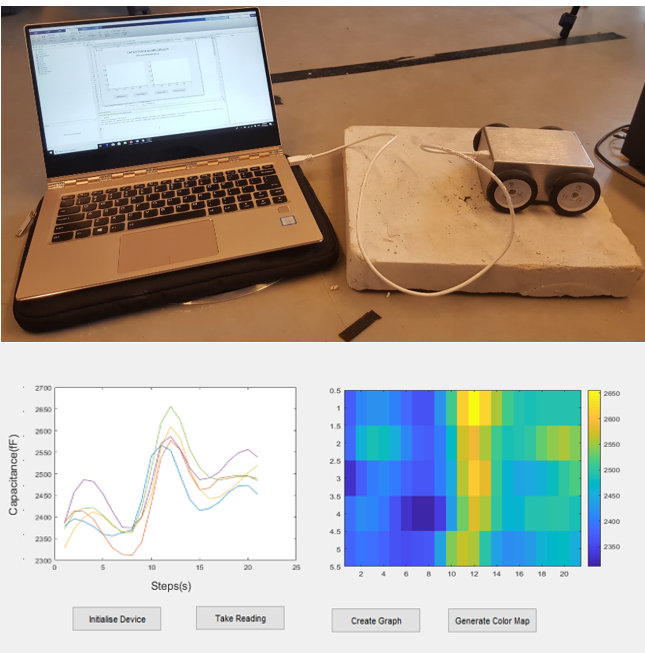}
    \caption{Scans are done by moving the sensor on the top surface of the test area. The Graphical User Interface (GUI)  displays the 2D image of the underlying subsurface.}
    \label{fig:sensor_setup}
\end{figure}

In recent years, use of non-conventional sensors that can be deployed using mobile robots  to evaluate building infrastructure has increased due to the reason that it can efficiently carry out timely inspection even in places that human cannot easily reach \cite{gunatilake2019real,Thiyagarajan2018RobustInfrastructures,wickramanayake2019frequency,giovanangelia2019design,Thiyagarajan2018RobustSewers}. Those robots need low cost,  lightweight sensors with real-time processing capability for non destructive evaluation of buildings. Often integrating an off-the-shelf sensor with robots is difficult due to cost, larger footprint and non-availability of raw data. 

To overcome these barriers, and with the intention of future integration of robotic deployment mechanism, this paper introduces a capacitive based sensor technology that can image the underline subsurface on flat building surfaces. The sensor is custom built with Finite Element Analysis (FEA) simulations, necessary hardware, and software components. Capacitive sensors have many advantages over other alternative sensing technologies. As the capacitance is largely dependent on the dielectric properties of the material, it can be used to detect underline structures made with a variety of building materials. The capacitive sensor is easy to construct and has a small footprint with low energy consumption. More importantly, capacitive sensors are more sensitive to the near-surface variations of the material. Those advantages make capacitive sensing an ideal choice for subsurface robotic imaging of building surfaces. 

The built capacitive sensor is shown in Fig. \ref{fig:sensor_setup}. It consists of a scan head that houses a capacitive sensor with its hardware, and software components that run on a computer. The raw data is sent to the computer via a USB cable,  and the Graphical User Interface (GUI) of the software process generates the   2D subsurface image as shown in Fig. \ref{fig:sensor_setup}. The scans head is manually moved by the operator on the test surface and multiple parallel scans are done to cover the area of interest. The wheel encoders attached to the scan head detect the movement of the sensor and corresponding capacitive data is recorded for each spatial resolution.

The rest of the paper describes the complete process we undertook to build the capacitive sensor for subsurface imaging, which includes FEA simulations, hardware, and software design and sensor validation through laboratory experiments. 

\section{The principle of Capacitance}
Capacitance is the ability of a body to store charge, and capacitive sensors work by measuring the change in capacitance caused by the influence of external objects. Thus measured capacitance variations can be used to detect changes of materials in the test surface. 

When the two copper electrodes are connected to the power source, it creates two oppositely charged plates. The positive charges and negative charges are attractive, so they move as close to each other as possible. Similar charges are repulsive, and they distribute themselves evenly in their respective plates to maximize the distance between them. As the charges build-up, an electric field is generated between the two electrodes. 
When an external object interferes the electric field between the two electrodes, it is polarised. This polarising causes energy to be transferred from the two electrodes to the body and the potential of the plates decreases in the process. As the electrodes are connected to the power source, more charge flows into the copper plates to regain the lost energy and the charge stored in the electrodes is increased resulting in a  change of capacitance values. 

As the sizes of the electrodes are increased, the capacitance and the percentage change of capacitance increases. This increases the sensing capability of the sensor. Another factor that affects the capacitance change is the distance between the electrodes \cite{yin2010non}. As the distance between the electrodes is decreased, the strength of the capacitance field increases. However, it decreases the percentage change of capacitance caused by the specimen being tested which in turn can decrease the sensitivity of the sensor \cite{yin2010non,laflamme2012soft,yin2013studies}.

When designing a capacitive sensor, one of the major challenges is maximizing the change in capacitance caused by the specimen being tested while minimizing the effect on the capacitance from neighboring objects. A  larger electrode is intended to provide constant coupling to the external interference whereas a  smaller electrode can accurately measure the change in capacitance caused by the specimen being tested with a better spatial resolution. It increases the sensitivity of the sensor and results in a larger percentage change of capacitance from the test specimen.

\section{Sensor Head Design}
FEA simulations were conducted on Ansys Maxwell, to decide the parameters that influence the capacitance and to optimize the design of the sensor head. It was also used to find the shape of the electrodes that generates the maximum penetration depth through the material.

\subsection{Simulation Setup}
\begin{figure}
    \centering
    \includegraphics[width=0.2\textwidth]{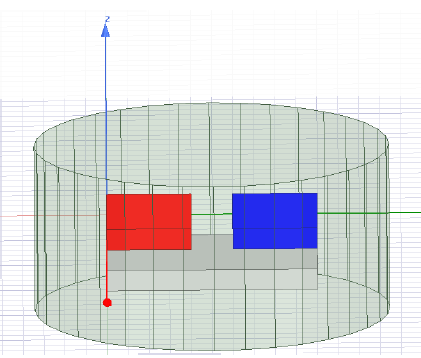}
    \caption{Simulation Setup in Ansys Maxwell}
    \label{fig:simulation_setup}
\end{figure}

Fig. \ref{fig:simulation_setup}  shows the basic setup for the simulations. The two copper electrodes are placed 1mm above the surface of the test sample. The resistivity of the test sample is set to $10^{12}$ ohm-mm (resistivity of dry concrete) and the setup is enclosed in a vacuum polyhedron. The positive electrode is supplied with a voltage of 5V and the negative electrode with a voltage of 0V. Uniform scale is used to compare the results of the simulations.

\subsection{Separation Distance Between the Electrodes}
\begin{figure}
     \centering
     \begin{subfigure}[b]{0.5\textwidth}
         \centering
         \includegraphics[width=0.7\textwidth]{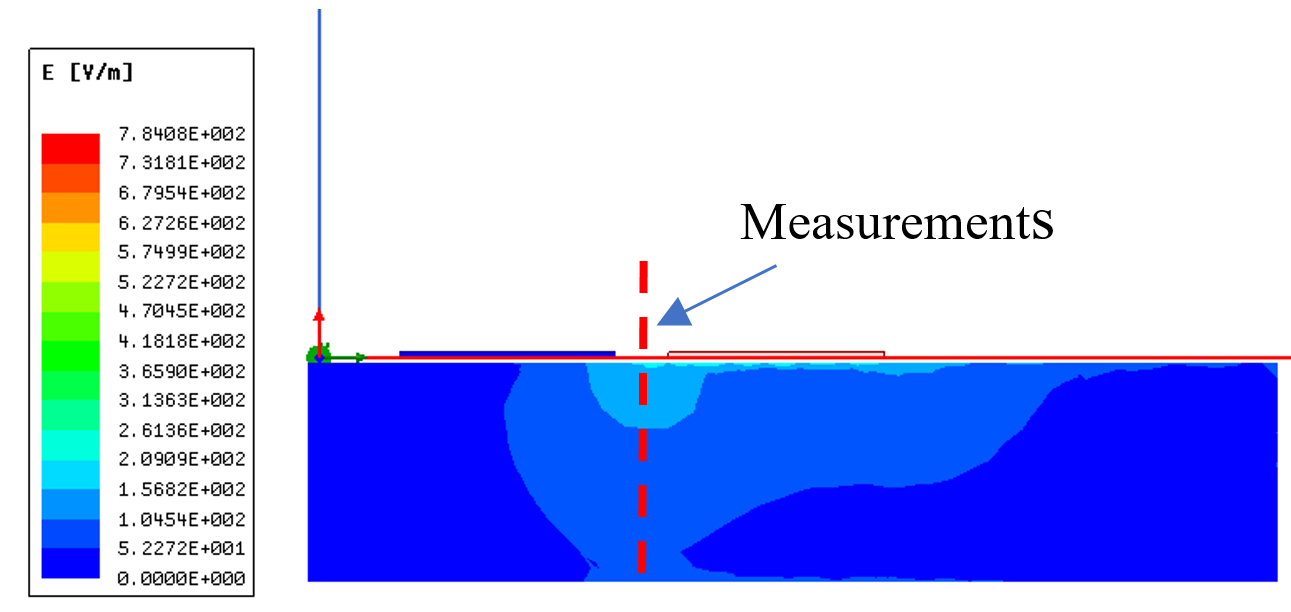}
         \caption{Ansys simulations by varying the electrode separation distance. Side view of the simulation is shown}
         \label{}
     \end{subfigure}
     \hfill
     \begin{subfigure}[b]{0.5\textwidth}
         \centering
         \includegraphics[width=0.7\textwidth]{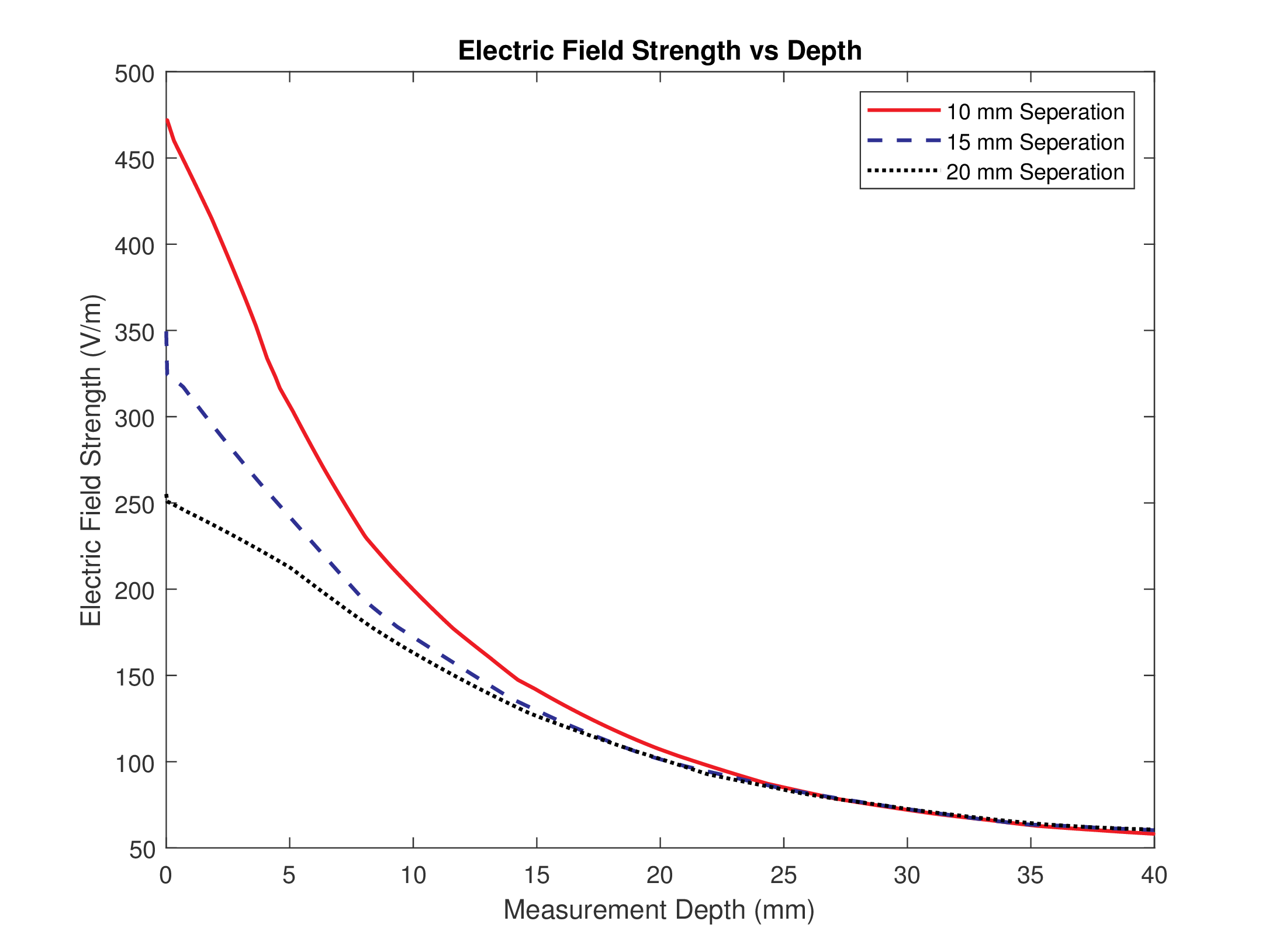}
         \caption{Electric Field Strength Vs Depth. Measurements were taken along a center line between the electrodes by varying the electrode separation distance}
         \label{}
     \end{subfigure}
    
        \caption{Simulations by varying electrode separation distance}
        \label{fig:simulations_electrode_seperation}
\end{figure}

Simulations were conducted by changing the distance between the two electrodes while keeping all the other parameters constant. The change in the penetration strength of the electric field generated was then analyzed. Measurements are taken along a centerline between the two electrodes, and the simulation results are shown in Fig. \ref{fig:simulations_electrode_seperation}. By comparing the results, we can see that as the distance between the two electrodes increase, the penetration strength of the electric field in the sample specimen decreases. Also, simulation results showed strong field lines between the two electrodes. This can be explained by Coulomb's law. As the distance between the electrodes increases, the electric field weakens according to the Coulomb's equation. The results are in favor of the context of the design as a small distance between the electrodes will produce the electric field with the highest penetration strength while reducing the size of the required electrodes. 

\subsection{Lift-off Distance of the Electrodes}
Simulations were conducted by changing the lift-off distance, the distance between the electrodes and the concrete, while keeping all the other parameters constant. The change in the penetration strength of the electric field was then analyzed.

\begin{figure}
     \centering
     \begin{subfigure}[b]{0.5\textwidth}
         \centering
         \includegraphics[width=0.67\textwidth]{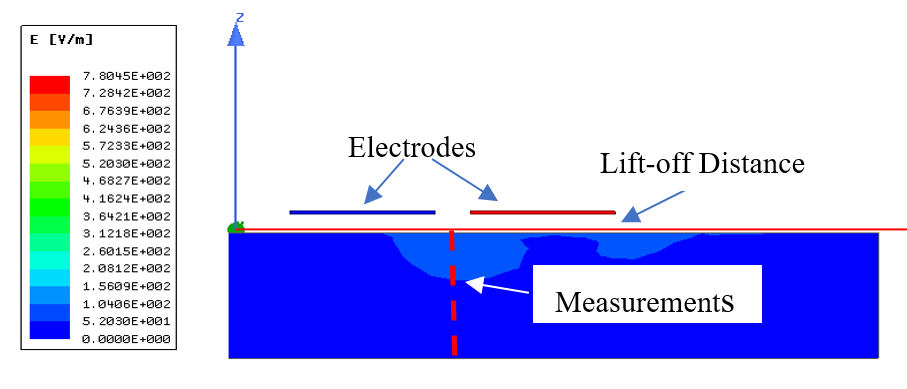}
         \caption{Ansys simulations by varying the electrode lift-off distance. The side view of the simulation is shown}
         \label{}
     \end{subfigure}
     \hfill
     \begin{subfigure}[b]{0.5\textwidth}
         \centering
         \includegraphics[width=0.67\textwidth]{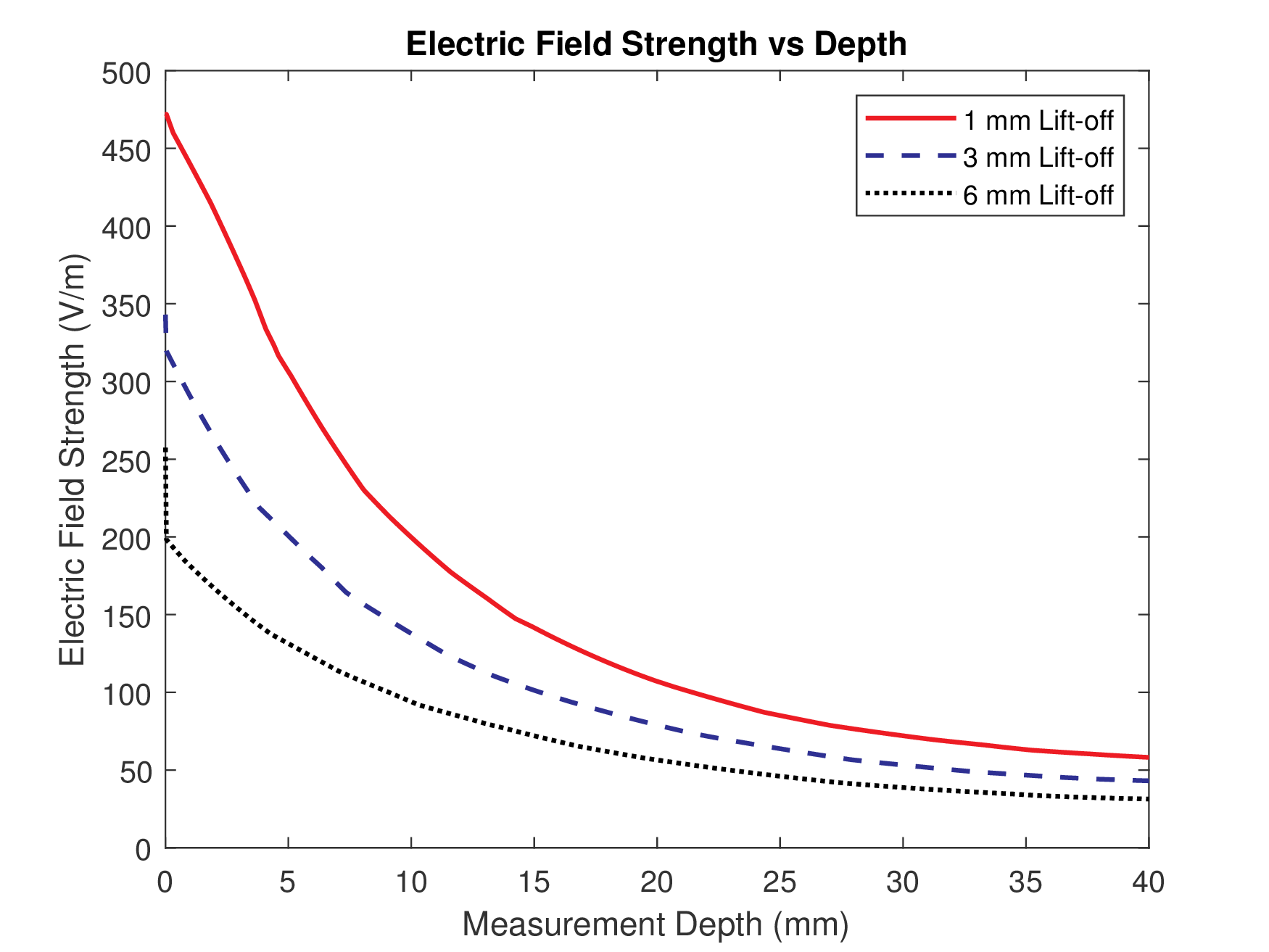}
         \caption{Electric Field Strength Vs Depth. Measurements were taken along a center line between the electrodes by varying the electrode lift-off distance}
         \label{}
     \end{subfigure}
    
        \caption{Simulations by varying electrode lift-off distance}
        \label{fig:simulation_lift_off}
\end{figure}

According to the simulation results shown in Fig. \ref{fig:simulation_lift_off},  as the distance between the concrete and the electrodes increases, the penetration strength of the electric field gradually get decreased which can be explained by the Coulomb's law like in the previous case. In the context of the design, closer the device is to the concrete surface, the penetration strength of the electric field generated will be higher. However, the device needs to be kept a few mm above the specimen on wheels to facilitate continuous measurements.

\subsection{Shape of the Electrode}
Due to the size constraint on the electrodes to make the device portable, different shapes of electrodes were simulated to find the optimum shape that generates the electric field with the strongest penetration strength. Common electrodes configurations such as comb, circular and triangular shape electrodes were simulated to decide the most suitable electrode shape. 

\begin{figure*}
     \centering
     \begin{subfigure}[b]{0.3\textwidth}
         \centering
         \includegraphics[width=\textwidth]{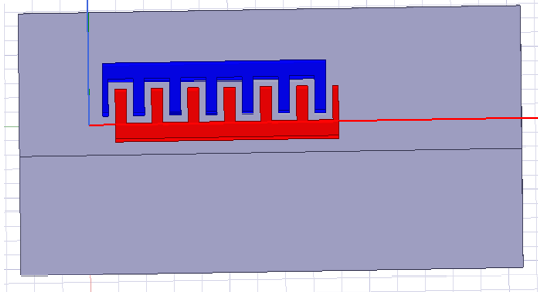}
         \caption{Comb shape electrode}
         \label{}
     \end{subfigure}
     \hfill
     \begin{subfigure}[b]{0.3\textwidth}
         \centering
         \includegraphics[width=\textwidth]{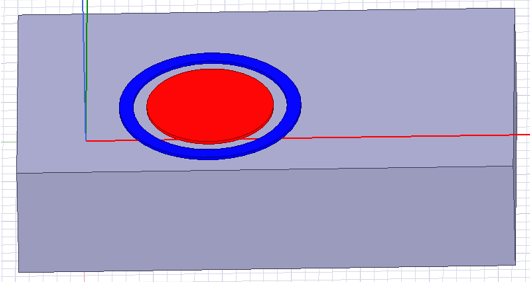}
         \caption{Circular shape electrode}
         \label{}
     \end{subfigure}
      \hfill
      \begin{subfigure}[b]{0.3\textwidth}
         \centering
         \includegraphics[width=\textwidth]{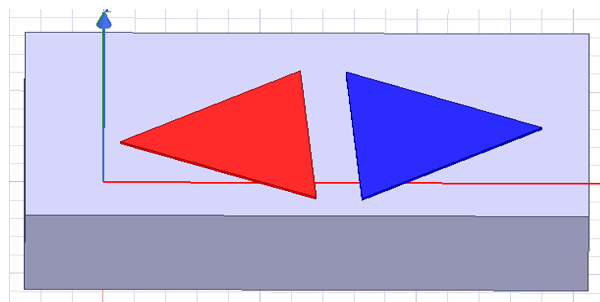}
         \caption{Triangular shape electrode}
         \label{}
     \end{subfigure}
      \hfill
    \begin{subfigure}[b]{0.3\textwidth}
         \centering
         \includegraphics[width=\textwidth]{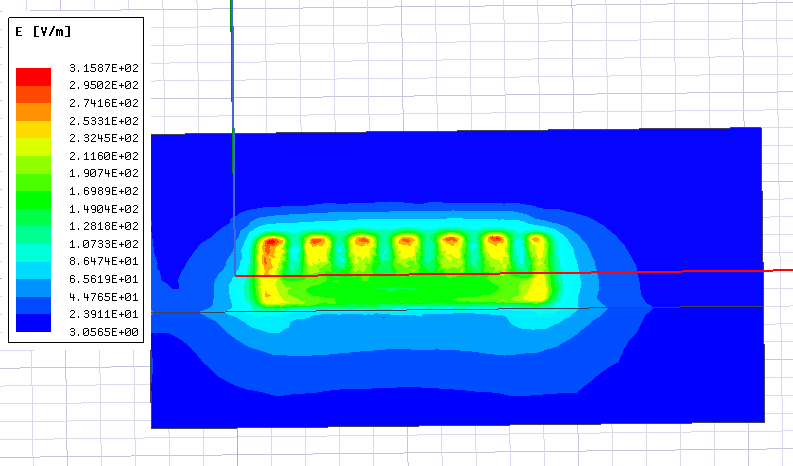}
         \caption{Ansys simulations of the comb shape electrode}
         \label{}
     \end{subfigure}
     \hfill
     \begin{subfigure}[b]{0.3\textwidth}
         \centering
         \includegraphics[width=\textwidth]{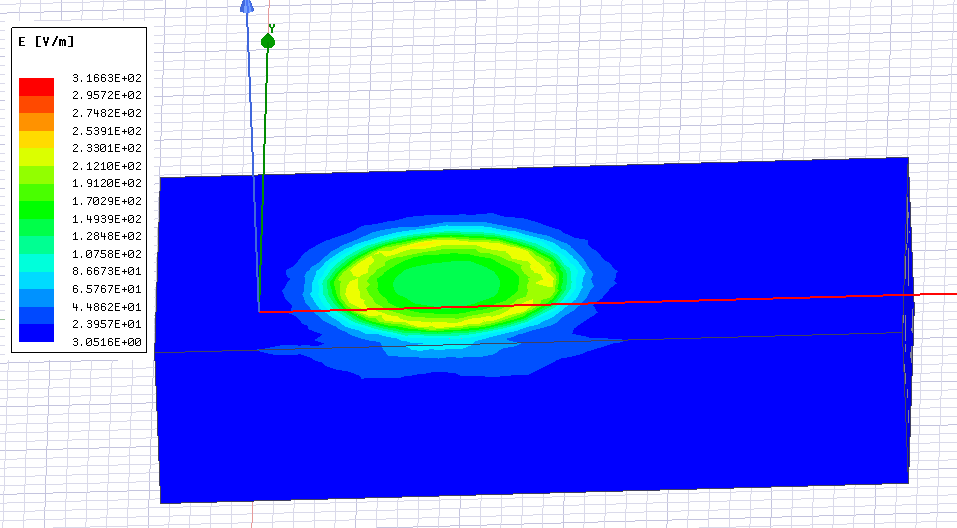}
         \caption{Ansys Simulation of the  circular shape electrode}
         \label{}
     \end{subfigure}
      \hfill
      \begin{subfigure}[b]{0.3\textwidth}
         \centering
         \includegraphics[width=\textwidth]{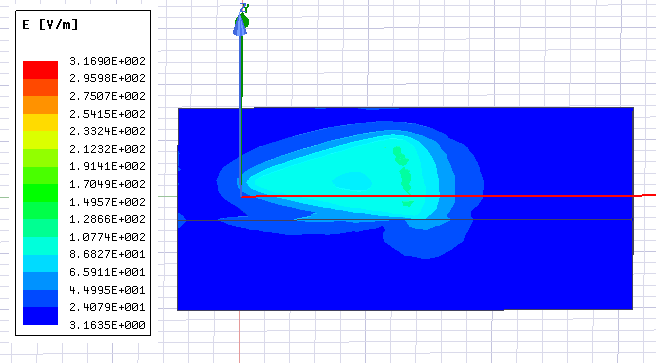}
         \caption{Ansys Simulation of the  triangular shape electrode}
         \label{}
     \end{subfigure}
      \hfill

        \caption{Simulations for different electrode shapes}
        \label{fig_electrode_shape}
\end{figure*}

According to the simulation results shown in Fig. \ref{fig_electrode_shape}, comb-shaped electrodes create the electric field with the strongest penetration strength. The second strongest field is generated by the circular electrodes. The penetration strength of the triangular-shaped electrode is very low compared to the other two shapes.

\section{Hardware Development}

\begin{figure}
    \centering
    \includegraphics[width=0.4\textwidth]{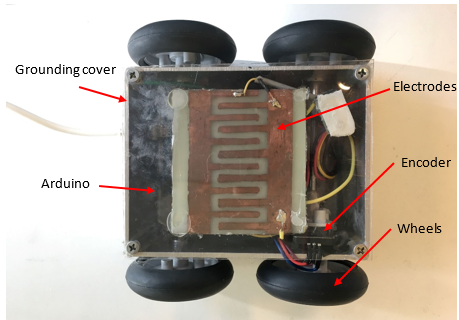}
    \caption{Hardware components of the sensor head.}
    \label{fig:sensor_componets}
\end{figure}

Fig. \ref{fig:sensor_componets}  shows the final prototype. The wheels are connected through an axle that passes through bearings to reduce the friction between the moving parts. A rotary encoder is used and attached to the rear axle to count wheel rotations. It is then connected to a micro-controller which is placed inside the metal enclosure. The sensor pad is then attached to a thin sheet of plexiglass and screwed to the bottom of the metal enclosure to protect the sensor head. 

\subsection{Sensor Head}
Based on the research and simulations, it was found that electrodes shaped like a comb generated the electric field with the highest penetration strength. This was used in the final prototype. A thin aluminum plate was attached to the backside of the sensor to work as an active shield to reduce the interference from the nearby objects on the sensor readings. 

\subsection{Sensor Electronic}

A high-resolution capacitance-to-digital converter from Texas Instruments (FDC 1004)  was used to read the capacitive values from the sensor head \cite{FC1004site}. It has four channels with a range of ±15 pF and can be programmed through the internal digitally programmable capacitor (CAPDAC) to handle capacitance up to 100 pF. The sensor also has a shield that can be used to reduce parasitic capacitance and the interference from the surrounding objects. 
Conductors are used at the backside of the electrodes and connected to the shield to reduce the effect of noise and stray capacitance on the results.

FDC 1004 implements a switched capacitor circuit to transfer charge from the sensor electrode to the sigma-delta analog to digital converter (ADC). The sensor uses a 25 kHz step waveform to charge the electrode and this charge is passed on to the sample hold circuit.  The resulting analog voltage is converted to a digital signal by the sigma-delta ADC which is then filtered according to the gain and offset assigned by the user.

\subsection{Shielding}
The major challenges with using capacitance for sensing are the effect of parasitic capacitance, environmental changes (humidity and temperature) and the effect from surrounding objects. FDC1004 has a shield that helps to reduce the effect of these factors on the result. The shield works by sending a signal as the sensor input signal voltage so that there is no potential difference between the sensor input signal and the shield. Any external interference will couple to the shield electrode with minimal interaction with the sensor electrode.

A shielded cable is used to connect the sensor to the electrodes and the shield of the cable is connected to the sensor shield. The sensor electronics were placed close to the electrodes to reduce the effect of parasitic capacitance between the input channel and the ground. A thin sheet of aluminum was attached to the rear end of the electrodes and connected to the shield to reduce the effect of external interference on the result.

\subsection{Wheel Encoder}

A rotary encoder is attached to the axle connecting the rear wheels of the device and used to keep track of its traveled distance. The microcontroller is programmed to take a reading from the capacitance sensor for every pulse generated by the rotary encoder. A hollow shaft rotary encoder, which generates 16 pulses for every revolution is used in the sensor. As the diameter of the wheel is 58.5mm, the device moves 183.78 mm for every wheel rotation,  resulting in measurements on every  11.5mm of sensor movement.

\subsection{Metal Enclosure}

One of the major challenges with using capacitance for sensing is the effect of external objects on the results. This can be considerably reduced by using a metal enclosure. The metal enclosure creates a Faraday Cage around the sensor and the reduces the effect of external charged objects on the results. As a charged object is brought near the metal enclosure, the charges in the metal enclosure rearrange itself to negate the effect of the external charged object on the components inside the enclosure. In the initial prototype, an aluminum enclosure of thickness 1mm and with dimensions of  116x90x60 ${mm^3}$ was used.

\section{Software Architecture}

The device is programmed to be controlled by a GUI running on the computer. The micro-controller inside the sensor enclosure measures the capacitance value for every step turn of the wheel encoder.  This data is sent through a serial link to the computer, where it gets analyzed and displayed. At the start of the program, the rotary encoder and the sensor are initialized. If a pulse is detected from the rotary encoder, the step number is increased, and a reading is taken from the sensor. The capacitance value is calculated using the reading from the sensor and calibrated using the  CAPDAC value. The reading from the sensor is analyzed and auto-calibrated if it is not within the readable range of the sensor.

\section{Experiments}
Figure \ref{fig:sensor_setup} shows the experimental setup for testing. The device is set up by connecting the sensor to the computer using a USB cable. A MATLAB based GUI is used to process the data and to generate the 2D subsurface image.

\subsection{Detecting Metal Rods Placed Behind a Plywood Sheet}

\begin{figure}
     \centering
     \begin{subfigure}[b]{0.5\textwidth}
         \centering
         \includegraphics[width=\textwidth]{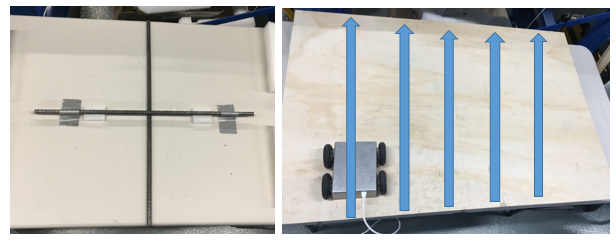}
         \caption{Two metal bars are  placed behind a plywood sheet and 5 vertical scans are done on the top surface  }
         \label{fig:experimental_metal_bar_setup}
     \end{subfigure}
     \vspace{12pt}
     \begin{subfigure}[b]{0.22\textwidth}
         \centering
         \includegraphics[width=\textwidth]{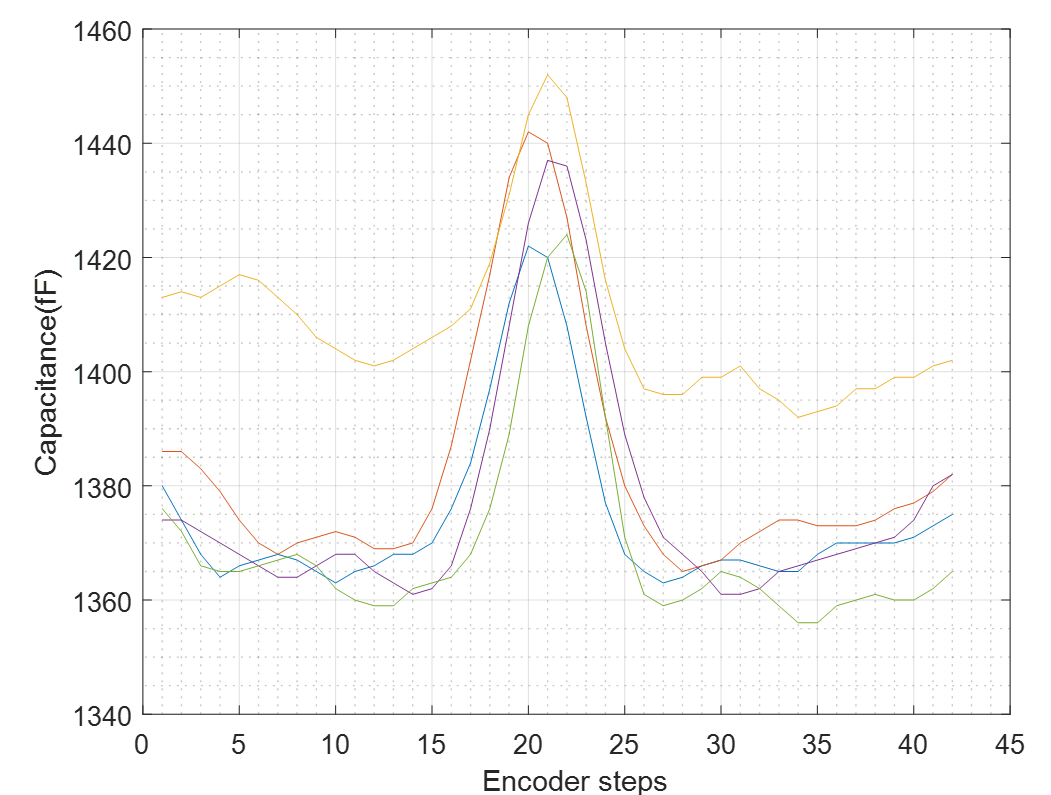}
         \caption{Capacitance values for each scan  }
         \label{fig:experimental_metal_bar_raw_data}
     \end{subfigure}
     \qquad
     \begin{subfigure}[b]{0.22\textwidth}
         \centering
         \includegraphics[width=\textwidth]{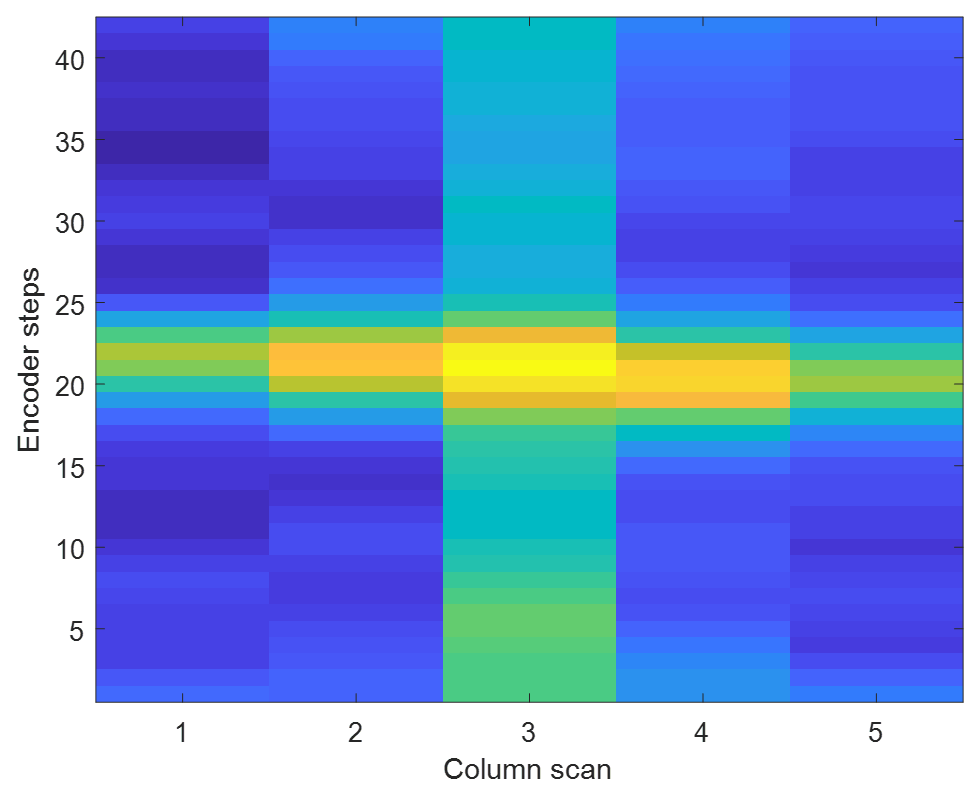}
         \caption{ 2D Subsurface image }
         \label{fig:experimental_metal_bar_2d_image}
     \end{subfigure}
        \caption{Detecting metal bars located behind a plywood sheet}
        \label{fig:experimental_metal_bar}
\end{figure}

To test the sensor for its ability to image the subsurface, two metal bars were placed behind 25mm plywood sheet vertically and horizontally. The sensor was then moved vertically as shown in Fig. \ref{fig:experimental_metal_bar_setup}. A total of five scans were done in the vertical direction. The generated 2D subsurface image is shown in Fig. \ref{fig:experimental_metal_bar_2d_image} and the raw capacitance values for each scan are shown in Fig. \ref{fig:experimental_metal_bar_raw_data}.
The 2D subsurface image clearly indicates the location of the metal bars and their orientations. High capacitive changes were evident when the scan head is on top of the metal bar. It is clear from these results the built sensor can locate the metal bars and their structure effectively, giving visual information about the subsurface behind the plywood sheet. 

\subsection{Detecting a Metal Bar Placed Behind a Concrete Sample}

\begin{figure}
     \centering
     \begin{subfigure}[b]{0.4\textwidth}
         \centering
         \includegraphics[width=\textwidth]{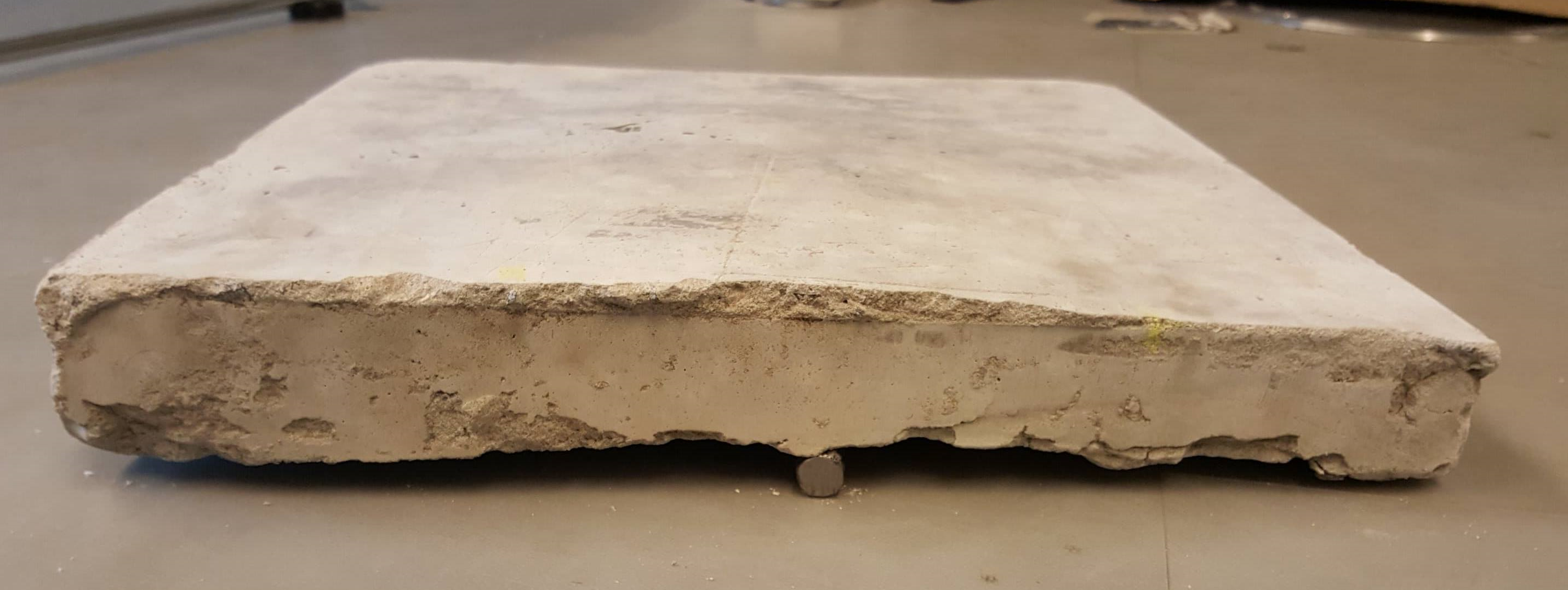}
         \caption{A metal bar is placed behind a 35 mm thick concrete samples. scans are done on the top surface, perpendicular direction to the bar}
         \label{}
     \end{subfigure}
     \vspace{12pt}
     \begin{subfigure}[b]{0.22\textwidth}
         \centering
         \includegraphics[width=\textwidth]{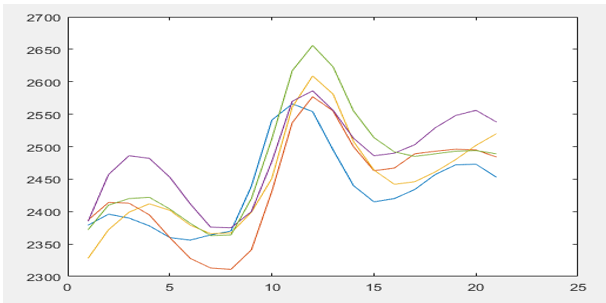}
         \caption{Capacitance values for each scan  }
         \label{}
     \end{subfigure}
     \qquad
     \begin{subfigure}[b]{0.22\textwidth}
         \centering
         \includegraphics[width=\textwidth]{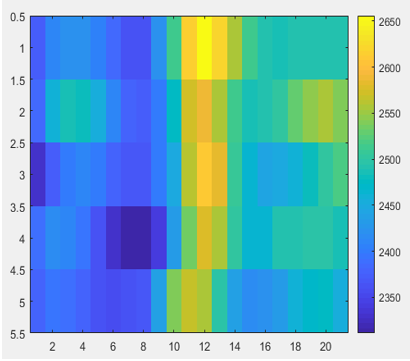}
         \caption{ 2D Subsurface image with clear indication of the metal bar location }
         \label{}
     \end{subfigure}
        \caption{Detecting metal bars located behind a concrete sample}
        \label{fig:experimental_metal_bar_concrete}
\end{figure}

Plywood and concrete are the most common building materials used in the construction industry.  As the device was able to accurately locate metal bars behind plywood surfaces, the device’s ability for locating metal bars behind concrete was tested. This is to understand the behavior of the sensor in different building materials. A Rebar is placed behind a 35mm thick concrete slab as shown in Fig. \ref{fig:experimental_metal_bar_concrete}, and the sensor was moved in a direction  perpendicular to  the rebar orientation. Five scans were done and the 2D subsurface image was generated using the GUI. The results are shown in Fig. \ref{fig:experimental_metal_bar_concrete}. 

It is clear that the capacitance values taken during the midway, where the metal rod is located, were considerably higher than the others. This is consistent with the findings from the previous tests done on plywood,  and an increase in capacitance above the metal bar location is clearly visible in the 2D subsurface image. This is a clear indication that our sensor can be used to detect metal bars and their structure both in concrete and plywood building materials effectively. 

\subsection{Detecting Wooden Studs Behind Walls}

\begin{figure}
     \centering
     \begin{subfigure}[b]{0.18\textwidth}
         \centering
         \includegraphics[width=\textwidth]{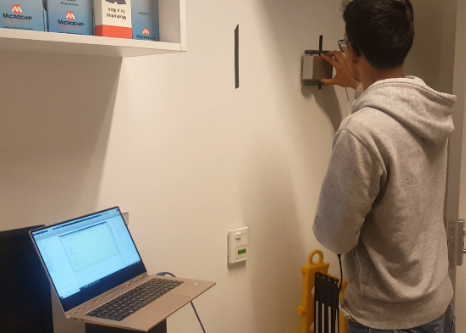}
         \caption{Scan are done on a wall with known stud location  }
         \label{}
     \end{subfigure}
     \qquad
     \begin{subfigure}[b]{0.22\textwidth}
         \centering
         \includegraphics[width=\textwidth]{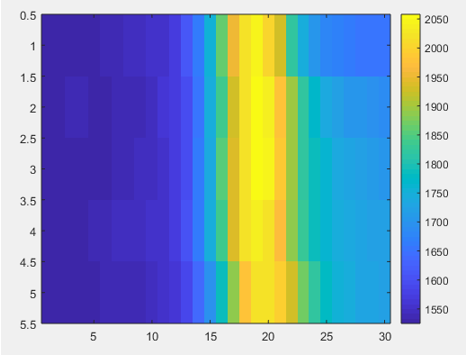}
         \caption{ 2D Subsurface image with  a clear indication of the stud orientation and location }
         \label{}
     \end{subfigure}
        \caption{Detecting wooden studs behind a wall}
        \label{fig:experimental_wooden_studs}
\end{figure}

Wooden studs are used heavily to build the structure of the buddings. Often these studs are hidden behind the wall and it is not easy to locate them. As such, we tested our sensor on a wall to its ability to locate the wooden studs and their orientations. The sensor was moved on a wall with a known stud location as shown in Fig. \ref{fig:experimental_wooden_studs}, and a total of five scans were done in the horizontal direction. The generated 2D subsurface image is shown in Fig. \ref{fig:experimental_wooden_studs}. As before,  the capacitance value increased considerably when the sensor is moving above the stud. When the five scans were combined together to foam the subsurface image the location and the orientation of the stud were clearly visible. This proves our sensor is highly usable in detecting studs and their orientations. 

\subsection{Detecting Both Studs and Metal Bars}

\begin{figure}
     \centering
     \begin{subfigure}[b]{0.35\textwidth}
         \centering
         \includegraphics[width=0.5\textwidth]{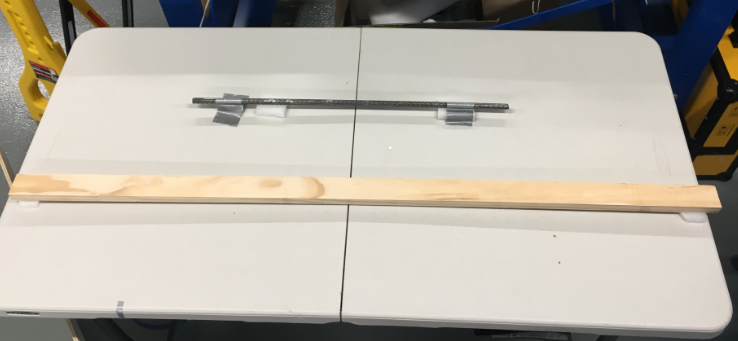}
         \caption{A metal bar and a wood bar are  placed behind a 25 mm plywood sheet. Scans were done on the top surface, perpendicular to  the bars }
         \label{fig:y equals x}
     \end{subfigure}
     \vspace{12pt}
     \begin{subfigure}[b]{0.22\textwidth}
         \centering
         \includegraphics[width=\textwidth]{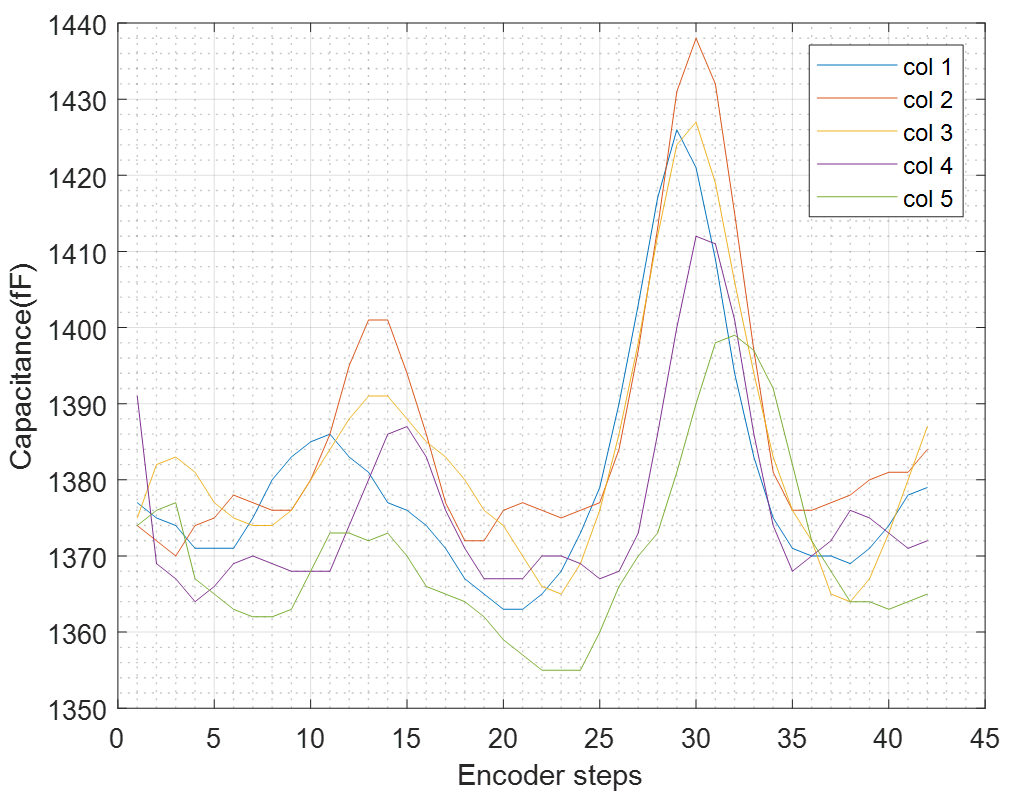}
         \caption{Capacitance values for each scan. Metal bar resulted in a higher peak than than the wooden bar }
         \label{}
     \end{subfigure}
     \qquad
     \begin{subfigure}[b]{0.22\textwidth}
         \centering
         \includegraphics[width=\textwidth]{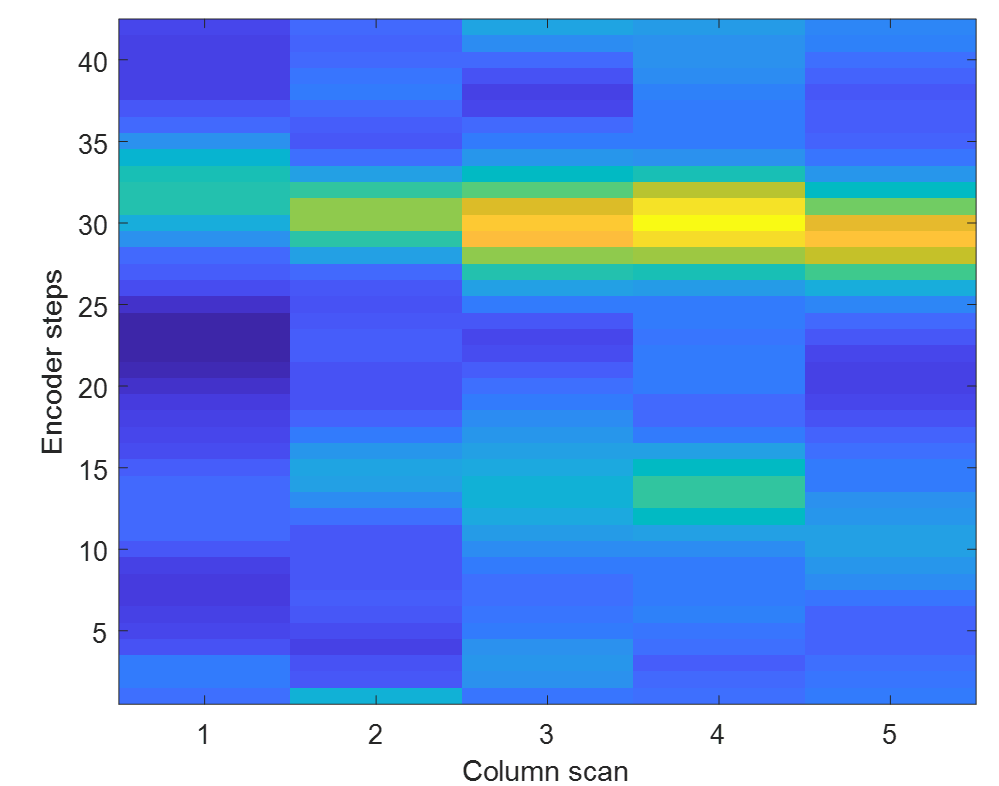}
         \caption{ 2D Subsurface image with  indication of  metal and wood bar locations}
         \label{}
     \end{subfigure}
        \caption{Detecting metal bars and wooden bars located behind a plywood sheet}
        \label{fig:experimental_metal_and_wood}
\end{figure}

To test the sensor's ability to differentiate metals bars and wooden studs, an experiment was done by placing a wooden stud and a metal bar behind a plywood sheet of 25mm thickness as shown in Fig. \ref{fig:experimental_metal_and_wood}. Five scans were done in the perpendicular direction to the metal and wood bars.  The scan results are shown in Fig. \ref{fig:experimental_metal_and_wood}. As can be seen from the 1D plot, a high capacitance peak is visible when the sensor is above the metal bar and a relatively small peak is visible when the sensor is above the wooden bar. The same observation is noticed in the 2D image as well. It is clear that from these results the capacitance value can be used to get an idea about the build material of studs hidden behind walls. With our sensor, high capacitance values indicate metal-like structures,  and low capacitance values indicate wooden like structures.

\section{Conclusions}
This paper introduced a novel capacitive sensor-based technology which can generate a 2D subsurface image to understand the underlying structure of building surfaces. FEA simulations were done to understand the optimum parameters for sensor head design. Hardware and software components are built to facilitate 2D subsurface imaging capability. Lab experiments are done to validate the sensor's usability, which resulted in encouraging outcomes. The sensor can be used successfully to reveal the underlying structure beneath building surfaces like walls and floors built with common building materials like wood and concrete. In the future, the machine learning model proposed in \cite{Thiyagarajan2018GaussianInfrastructure} can be incorporated into the system for automatic localization of underlying structures.

\bibliographystyle{IEEEtran}
\bibliography{conference}

\end{document}